\documentclass[11pt,twoside]{article}  % Leave intact
\usepackage{asp2006}
\usepackage{adassconf}

\begin{document}   % Leave intact

%-----------------------------------------------------------------------
%			    Paper ID Code
%-----------------------------------------------------------------------
% Enter the proper paper identification code.  The ID code for your paper 
% is the session number associated with your presentation as published 
% in the official conference proceedings.  You can find this number by 
% locating your abstract in the printed proceedings that you received 
% at the meeting, or on-line at the conference web site.
%
% This identifier will not appear in your paper; however, it allows different
% papers in the proceedings to cross-reference each other.  Note that
% you should only have one \paperID, and it should not include a
% trailing period.

\paperID{P27}

%-----------------------------------------------------------------------
%		            Paper Title 
%-----------------------------------------------------------------------
% Enter the title of the paper.
%
% EXAMPLE: \title{A Breakthrough in Astronomical Software Development}

\title{GPU-Based Volume Rendering of Noisy Multi-Spectral Astronomical Data}
       
%-----------------------------------------------------------------------
%          Short Title & Author list for page headers
%-----------------------------------------------------------------------
% Please supply the author list and the title (abbreviated if necessary) as 
% arguments to \markboth.
%
% The author last names for the page header must appear in one of 
% these formats:
%
% EXAMPLES:
%     LASTNAME
%     LASTNAME1 and LASTNAME2
%     LASTNAME1, LASTNAME2, and LASTNAME3
%     LASTNAME et al.
%
% Use the "et al." form in the case of four or more authors.
%
% If the title is too long to fit in the header, shorten it: 
%
% EXAMPLE: change
%    Rapid Development for Distributed Computing, with Implications for the Virtual Observatory
% to:
%    Rapid Development for Distributed Computing

\markboth{Hassan, Fluke, and Barnes}{GPU-Based Volume Rendering}

%-----------------------------------------------------------------------
%		          Authors of Paper
%-----------------------------------------------------------------------
% Enter the authors followed by their affiliations.  The \author and
% \affil commands may appear multiple times as necessary.  List each
% author by giving the first name or initials first followed by the
% last name. Do not include street addresses and postal codes, but 
% do include the country name or abbreviation. 
%
% If the list of authors is lengthy and there are several institutional 
% affiliations, you can save space by using the \altaffilmark and \altaffiltext 
% commands in place of the \affil command.

\author{Amr H.\ Hassan, Christopher J.\ Fluke, and David G.\ Barnes}
\affil{Centre for Astrophysics \& Supercomputing, Swinburne University
of Technology, Hawthorn, Victoria, Australia}

% Notice that some of these authors have alternate affiliations, which
% are identified by the \altaffilmark after each name.  The actual alternate
% affiliation information is typeset in footnotes at the bottom of the
% first page, and the text itself is specified in \altaffiltext commands.
% There is a separate \altaffiltext for each alternate affiliation
% indicated above.

%-----------------------------------------------------------------------
%			 Contact Information
%-----------------------------------------------------------------------
% This information will not appear in the paper but will be used by
% the editors in case you need to be contacted concerning your
% submission.  Enter your name as the contact along with your email
% address.

\contact{Amr Hassan}
\email{ahassan@swin.edu.au}

%-----------------------------------------------------------------------
%		      Author Index Specification
%-----------------------------------------------------------------------
% Specify how each author name should appear in the author index.  The 
% \paindex{ } should be used to indicate the primary author, and the
% \aindex for all other co-authors.  You MUST use the following syntax: 
%
%    \aindex{LASTNAME, F.~M.}
% 
% where F is the first initial and M is the second initial (if used). Please 
% ensure that there are no extraneous spaces anywhere within the command 
% argument. This guarantees that authors that appear in multiple papers
% will appear only once in the author index. Authors must be listed in the order
% of the \paindex and \aindex commmands.

\paindex{Hassan, A. H.}
\aindex{Fluke, C. J.}
\aindex{Barnes, D. G.}

%-----------------------------------------------------------------------
%			Subject Index keywords
%-----------------------------------------------------------------------
% Enter up to 6 keywords that are relevant to the topic of your paper.  These 
% will NOT be printed as part of your paper; however, they will guide the creation 
% of the subject index for the proceedings.  Please use entries from the
% standard list where possible, which can be found in the index for the 
% ADASS XVI proceedings. Separate topics from sub-topics with an exclamation 
% point (!). 


% We reset the footnote counter for the hyperlink since it does not
% appear to recognize the previous 3 footnotes generated from the
% altaffilmarks.  

\setcounter{footnote}{3}

%-----------------------------------------------------------------------
%			       Abstract
%-----------------------------------------------------------------------
% Type abstract in the space below.  Consult the User Guide and Latex
% Information file for a list of supported macros (e.g. for typesetting 
% special symbols). Do not leave a blank line between \begin{abstract} 
% and the start of your text.

\begin{abstract}          % Leave intact
Traditional analysis techniques may not be sufficient for astronomers to make the best use of the data sets that current and future instruments, such as the Square Kilometre Array and its Pathfinders, will produce. By utilizing the incredible pattern-recognition ability of the human mind, scientific visualization provides an excellent opportunity for astronomers to gain valuable new insight and understanding of their data, particularly when used interactively in 3D. The goal of our work is to establish the feasibility of a real-time 3D monitoring system for data going into the Australian SKA Pathfinder archive.

Based on CUDA, an increasingly popular development tool, our work utilizes the massively parallel architecture of modern graphics processing units (GPUs) to provide astronomers with an interactive 3D volume rendering for multi-spectral data sets. Unlike other approaches, we are targeting real time interactive visualization of datasets larger than GPU memory while giving special attention to data with low signal to noise ratio - two critical aspects for astronomy that are missing from most existing scientific visualization software packages. Our framework enables the astronomer to interact with the geometrical representation of the data, and to control the volume rendering process to generate a better representation of their datasets. 
\end{abstract}

%-----------------------------------------------------------------------
%			      Main Body
%-----------------------------------------------------------------------
% Place the text for the main body of the paper here.  You should use
% the \section command to label the various sections; use of
% \subsection is optional.  Significant words in section titles should
% be capitalized.  Sections and subsections will be numbered
% automatically. 

\section{Introduction}

Next-generation astronomy research facilities will generate new challenges for data storage, access, analysis and system monitoring, bringing astronomy into the Petascale Data Era. But even today, astronomical knowledge is not growing at the same rate as the data. Scientific visualization is a fundamental, enabling technology for knowledge discovery. Despite recent progress, many current astronomy visualization approaches will be seriously challenged by, or are completely incompatible with, the Petascale Data Era.  With an emphasis on developing new approaches compatible with data from the Square Kilometer Array and its Pathfinders, the goal of this work is to advance the field of astrophysical visualization in preparation for the Petascale Era. 

\section{Challenges and Design Objectives}

The main goal of this work is to enable astronomers to visualize large spectral data cubes (e.g. at least 1 TB in size) of the size that will be generated from \htmladdnormallinkfoot{the Australian SKA Pathfinder (ASKAP)}{http://www.atnf.csiro.au/projects/askap/}. Multispectral data can be considered as a 4D data volume where three dimensions are associated to position allocation (two dimensions for the spatial position in sky coordinates and one dimension for the wavelength or frequency - both of which are related to the line of sight velocity) and one dimension for the flux density. The data cube can be considered as a stack of images where each image presents a sky portion over a small frequency range ($\Delta$$\nu$).  To achieve this target we designed a framework that utilizes the latest available hardware technologies combined with the latest software infrastructure. The main design objectives and challenges for this framework were:
\begin{enumerate}
\item Being scalable enough to offer better visualization outcomes when more hardware is available;
\item Support heterogeneous computing systems;
\item Using off-the-shelf hardware solution, such as graphics processing units (GPUs);
\item Requires minimum user intervention to reduce the time needed by astronomers to run and install the system;
\item Being capable to handle tiled display systems to provide the user with high resolution output; and
\item Handle data sizes exceeding current single machine memory limits.
\end{enumerate}
To achieve these objectives, the following design decisions were taken:
\begin{enumerate}
\item  \textbf {Using ray-casting volume rendering as our visualization technique.}
Although being computationally intensive and relatively hard to implement, volume rendering is an important visualization tool in our case because it gives the user a global picture of the data cube; it does not need the user to know in advance what s/he is searching for; and provides visuals that are easy to understand. Also, using ray-casting will support our framwork to achieve a high resolution output without any visual artifacts (Schwarz 2007).
\item \textbf{Build a framework based on a distributed GPU architecture}.
With a theoretical peak performance greater than 4 TFLOP/S ($10^{12}$ floating point operations per second) on hardware such as NVIDIA's Tesla, general purpose computing on graphics processing units offers a more effective and cheaper parallel architecture than existing multi-core processors. 
\end{enumerate}
\section{Distributed GPU Framework}
As shown in Figure 1, our distributed GPU framework combines the processing power of multiple GPU nodes to speed-up and enhance the spectral  cube visualization process.  This framework uses \htmladdnormallinkfoot{Message Passing Interface}{http://www.mcs.anl.gov/research/projects/mpi/} (MPI), multi-threading, and \htmladdnormallinkfoot{the Compute Unified Device Architecture}{http://www.nvidia.com/object/cuda_what_is.html} (CUDA) framework from NVIDIA to allow many GPUs to work on the same problem. 
\begin{figure}
\plotone{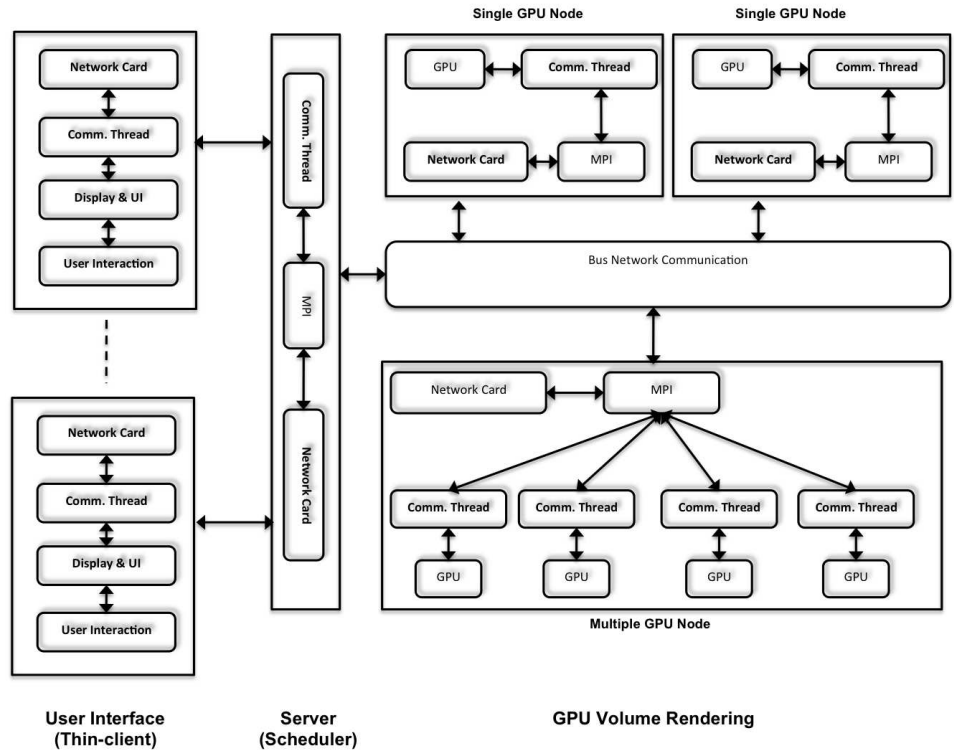}
\caption{Schematic diagram showing the framework's main hardware components. Each node is connected to the other nodes through a standard network interface which is managed through a seperate thread. The communication threads serve as a software communication channels between kernels on different GPUs. The scheduler server maps the current available tasks to the available computing resources. The scheduler module assumes that the resources are heterogenous and assign the computational tasks to each of them according to their computing capabilities. The final rendering result is presented to the user through a thin-client which translates the user interaction into rendering commands to the scheduler server.}
\label{fig:P27_1}
\end{figure}

\subsection{Framework Modules}
The main components of our proposed framework are partitioned based on their main functionality into: 
\begin{enumerate}
\item Scheduling module: responsible for managing and controlling the overall process;
\item GPU module: includes different execution kernels and utilizes the parallelization paradigm of the GPUs. We chose NVIDIA's CUDA library because it has a C like syntax, it is easy to use, and covers all the general purpose GPU computing functionality we require;  
\item Scene integration module: responsible for generating the final display output by combining the output of the contributing GPUs; and
\item I/O and User interaction modules: support the user's interactivity and change the output according to user input.
\end{enumerate}

The process of generating a single volume rendering view of the spectral line cube goes through the following processes:
\begin{enumerate}
\item The spectral data cube is partitioned into a set of smaller  sub-cubes; 
\item The scheduler module assigns these sub-cubes to the processing nodes/GPUs;
\item Each GPU  applies ray-tracing volume rendering to map each output pixel into a color plus alpha transparency channel (RGBA) based on a pre-selected transfer function;
\item The process of volume rendering produces N images with the same resolution as the output screen(s);
\item The scene integration module reapplies the same transfer function to these images to combine them into the final output; and
\item The final output is directed to the output display(s). 

\end{enumerate}
\section{Conclusion}
The focus of this work is to improve multi-spectral data visualization to cope with the vast amount of data to be produced by ASKAP and similar facilities. By employing GPUs combined with distributed processing, we are aiming to implement a scalable system capable of interactively visualizing ``greater than memory" astrophysical datasets.

\end{document}